\documentclass[12pt,preprint]{aastex}

\newcommand{\myemail}{chavarri@astrosen.unam.mx}

\slugcomment{Submitted to ``New Astronomy" 2013}

\shorttitle{PMS and ZAMS stars associated with LDN~1655}
\shortauthors{Chavarr\'{\i}a-K et al.}

\begin{document}

\title{PMS and ZAMS stars associated with the dark cloud LDN~1655}

\author{C. Chavarr\'{\i}a-K\altaffilmark{1,2,3}, M. A. Moreno-Corral\altaffilmark{1,2} 
and E. de Lara\altaffilmark{2}}
\affil{Instituto de Astronom\'{\i}a, Universidad Nacional Aut\'onoma de M\'exico, 
Ensenada B.C. 22860 (M\'{e}xico)}

\and 

\author{E. de la Fuente}  
\affil{Departamento de F\'{\i}sica CUCEI, 
Universidad. de Guadalajara, Guadalajara-Jalisco (M\'{e}xico)}

\altaffiltext{1}{Visiting Astronomer, Observatorio Astron\'omico Nacional (SPMO),
operated by the Instituto de Astronom\'{\i}a of the Universidad Nacional Aut\'onoma de M\'exico (IAUNAM).} 

\altaffiltext{2}{Instituto de Astronom\'{\i}a, Universidad Nacional Aut\'onoma de 
M\'exico/Campus Ensenada B.C.} 

\altaffiltext{3}{corresponding author, \myemail}

\begin{abstract}
We give results of a low-resolution optical spectroscopic study of the bright 
nebulous stars conforming the tight stellar trapezium embedded in IRAS06548-0815, 
of the exciting star of IRAS06547-0810, and of the stars associated with 
reflection nebulae (R- or N-stars) NJ065703.0-081421, NJ065714.1-081016,IRAS06548-0815~D 
and \mbox{IRAS06548-0815~G}, all objects scarcely observed and apparently pertaining 
to the dark cloud LDN~1655. Our results given here combined with the corresponding 
{\it 2MASS} near infrared photometry enables us to estimate a photometric distance 
to LDN1655 of 1.9$\pm 0.3$ kpc and to locate the trapezium stars on the 
($log \: L_*/L_\odot, log \: T_{eff}$) or HR diagram in an attempt to disclose about 
their true nature. 

The spectroscopy of \objectname{IRAS06548-0815} revealed two classic T Tauri, two 
Herbig Ae/Be stars, four weak-line T Tauri stars, and two probable lithium-rich 
stars. Additionally, we found that the R-star \objectname{IRAS06547-0810} is 
excited by a single B1(V) star, that \objectname{IRAS06548-0815}~D is excited by 
a SpT. B2/B3(V) star, NJ065703.0-081421 and NJ065714.1-081016 are excited by a 
B3/B4(V) emission star and an A6(V) star, respectively. The brightest near 
infrared source of the trapezium, IRS1, has an infrared luminosity comparable to 
a highly reddened O8(V) star (A$_V\approx 29^m$). The second brightest near 
infrared source of the trapezium, \objectname{IRAS06548-0815}~C is, optically, a 
single classical T Tauri star of spectral type K4/K5, but is a resolved binary 
in the {\sl 2MASS} $K_s$ band, both components being of about 
the same brightness. 
\end{abstract}


%
\section{Introduction} 
The purpose of our spectroscopic survey of selected objects associated with the 
dark cloud LDN1655 (Lynds 1962) 
is threefold: i) to detect its concomitant pre-main sequence stars, if any (i.e. 
to our knowledge, no such young stars have been reported in the literature), mainly 
the stars associated with IRAS06548-0815, ii) to derive a photometric distance to 
the cloud LDN1655 on hand of its associated (young blue) nebulous stars and iii) 
to locate its presumed young stars in the Hertzprung-Russell Diagram. On the 
following, we give a brief review of previous findings of the region. 

A quick inspection of the red POSS image of the region revealed a diffuse HII 
region obscured by a ``hammer-shaped'' dust lane in its foreground, the 
latter containing \object{IRAS06548-0815} and \object{IRAS06547-0810}, among other 
infrared sources and stars associated with reflection nebulosity (N- or 
R-stars, see Figure~1). Star \objectname{BD-07$\arcdeg$1642} ($V = 6\fm 3$,  
an A2V star at a distance $D\sim 170\;$pc, this paper) is the apparent brightest 
optical object of the region shown in Figure~1, located at its top-center section. 
The infrared source \object{IRAS06548-0815} stands out from the other sources 
of the region because of its mid infrared colors and morphology: it is a compact 
Orion-like trapezium\footnote{The distance to any two stars of the trapezium is 
comparable to its size.} entangled in reflection nebulosity (Magnier et al. 1999). 
The trapezium contains at least five nebulous stars, reminiscent of other similar 
very young stellar ensembles studied  by our group (e.g. Chavarr\'{\i}a-K et al. 
1987, 1989, 2005; Moreno-Corral et al. 1993, 2002, 2006; see also Gyulbudaghian 1995),
its stellar constituents could be medium and low-mass young stars. In support of this and 
considering that such clusters are loosely bound and hence subject to disruption by 
gravitational pull, the tightness of IRAS06548-0815 stellar aggregate (angular and 
linear di\'ameter $\phi \lesssim 1'$, $d \lesssim 0.3$ pc, respectively, assuming a 
distance to LDN1655 $D = 1.0$ kpc, see $\S 4$) signalizes 
us that gravity has not had the time to disperse it. In addition,  T Tauri 
stars are known to drift from their birth places typically with $v_d \simeq 1 - 
2\;km\:s^{-1}$ (e.g. Herbig 1977). 
Thus, the cluster crossing time for the case 
of IRAS06548-0815 is  $\sim 6\times 10^5$ yr (i.e. about the time required 
by an adiabatic perturbation to cross the  primitive knot), which is to be 
considered approximately  the age of the complex. Yet no report about the 
nature of its constituyents was found in the literature by us, except for 
two failed studies in radio wavelengths searching for massive (young) star 
formation (Codella et al. 1995; Magnier et al. 1999). 
Moreover, a closer inspection of 
the {\sl 2MASS} near-infrared (NIR) images of IRAS06548-0815 revealed us a 
tight spherical cluster of $\gtrsim 22$ stars tightly packed into a small 
volume (stellar density $n_{star} \backsimeq 1556\; $pc$^{-3}$, see Figure~2), 
remembrancing 
other (bright H$\alpha$) knots of known youth. 

Little is known about the region. Besides the information given in the  Dark Nebulae Catalog (LDN, 
Lynds  1962) 
and the IRAS Point Source Catalogue (Neugebauer et al. 1988),
only other five references covering the time span from 1850 to 2011 were retrieved from 
{the \sl ADS}, namely Maddalena et al. (1986), Codella et al. (1995), 
Hilton \& Lahulla (1995), Magnier et al. (1999) and Magakian (2003),  
but non of them discuss the nature of the associated nebulous stars. 

Regarding IRAS06547-0810, besides the Dark Cloud Nebulae and the IRAS catalog, 
no other references  were retrieved by SIMBAD. 
Only a single star is seen to be associated with the source in both POSS (blue 
\& red) prints, and in the off-set guider monitors of the 2.1 and 0.84 m telescopes, as 
well as in the {\it 2MASS} infrared images. From its infrared colors, the infrared source 
seems a normal reddened blue star and is located about half way 
on the dust-lane (i.e. LDN~1655) joining IRAS06548-0815 with \mbox{BD-07$^\circ 1642$} 
(c.f. Figure~1). Spectroscopy of IRAS06547-0810, together with that of other R- or N-stars in 
the region and their corresponding near infrared photometry enables us to directly estimate 
the distance to LDN1655, which has not been done previously. 

\begin{figure*}
\epsscale{1.0} 
\plotone{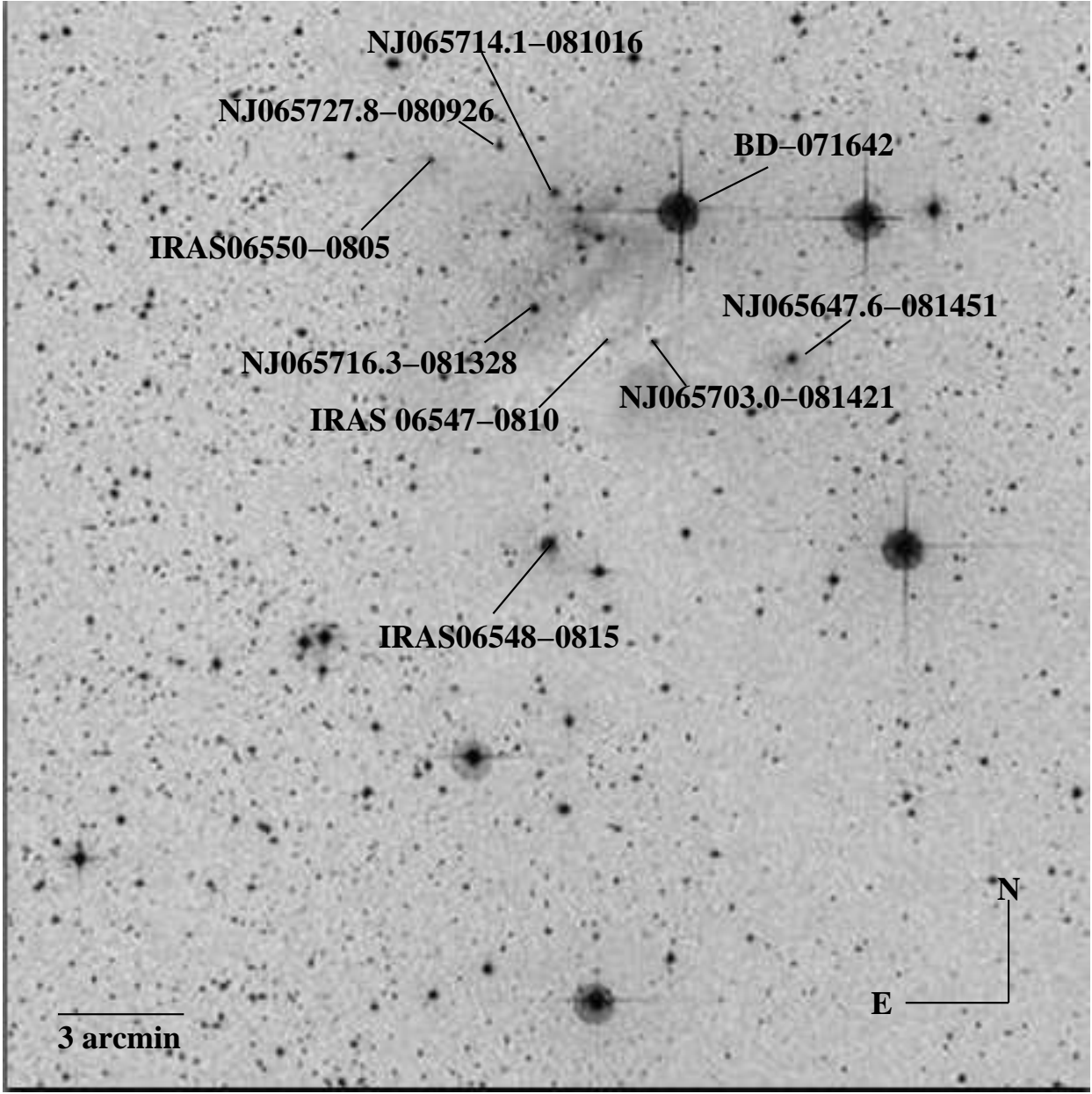}
\caption{A $30' \times 30'$ image taken from the Digital POSS with LDN~1655 at its 
     center, and where the stars and IRAS regions discussed herein this work are labeled.}  
\end{figure*} 
%

This paper is organized as follows:  $\S2$ describes the observations and reduction
techniques; $\S3$ contains the spectroscopic results of the optically bright stars 
associated with IRAS06548-0815, and of the (single) stars exciting IRAS06547-0810, 
NJ065703.0-081421 and  NJ065714.1-081016; in $\S4$ we derive a distance to LDN~1655, 
based on our spectroscopic results and the {\sl 2MASS} photometry; in $\S5$ we locate 
the bright stars associated with IRAS06548-0815 and LDN~1655 in the 
Hertzprung-Russell diagram; in $\S6$ we discuss the expected errors; and finally, in 
$\S7$ we discuss our results and give our conclusions. 

%
\begin{figure}
\epsscale{1.0}
\plotone{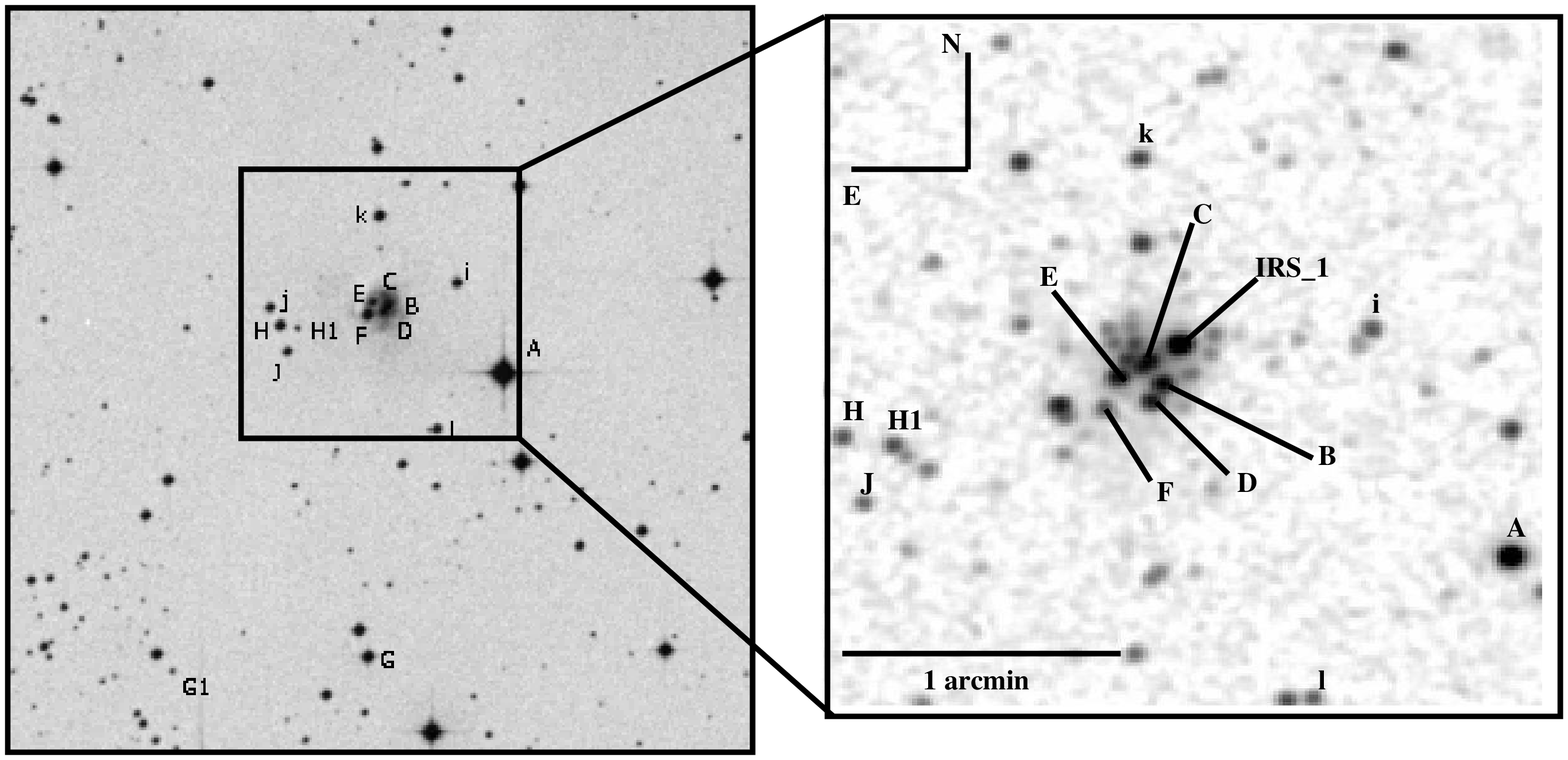}
\caption{Blow up of the central region of Figure~1 containing IRAS06548-0815 and
    showing the star designations used throughout this work. The right-side panel 
    shows the near infrared image in the {\sl 2MASS} $K_s$ band of IRAS06548-0815. 
    Same Sky orientations as in Figure~1. } 
\end{figure} %
%
\section{Observations and data reduction}  
\subsection{The observations}  
\label{sec:obs} 

In January 2006 we acquired low-resolution spectra ($\lambda/\Delta \lambda \simeq 
1600$ at H$\alpha$) of the brightest stars associated with IRAS06548-0815. We used 
the {\sl Italian} Boller \& Chivens spectrograph attached to the f/7.5 Cassegrain 
focus of the 2.1-m telescope of the Sierra San Pedro M\'artir National Astronomical 
Observatory (SPMO), with the SITE~\#3 CCD light detector. a $1024 \times 1024 \: 
pixel^2$ array with pixel size $= 24 \times 24\;\mu\:m^2$, and a CuHeNeAr arc lamp to 
calibrate in wavelength the dispersion axis. The detector was masked for our 
convenience to a width of 275 pixel perpendicular to the dispersion or right 
ascension axis. The plate scale is 0.347 "/pix. This CCD detector is linear over a 
wide dynamical 
range\footnote{more details in {\tt www.astrossp.unam.mx/Instruments/ccds/ccdcal/}}, 
with no significant fringing\ even by $\lambda = 8000$ \AA. The slit had an 
effective aperture of 150 $\mu$m (or $2\farcs1$, sky projected). A diffraction 
grating with a 600 lines mm$^{-1}$ ruling and blazed at $13^\circ$ was used in the 
first order for the observations. The grid angle was fixed at $11^\circ 30'$, 
optimizing the spectral resolution by the H$\alpha$ region (spectral resolution 
= 3.9 \AA~). The central wavelength of the spectrograms was $ 6260$ \AA ~and the 
spectral range covered was $5100$\AA$\leq \lambda \leq 7200$\AA, with a dispersion 
of 2.07 \AA~$pixel^{-1}$, but since the detector's surface normal was tilted with 
respect to the optical axis in the dispersion direction, the spectral resolution 
degraded accordingly and the spectrograms became noiser and wider at wavelengths 
shorter than about 5500 \AA, concentrating our analysis at longer wavelengths than 
this. The spectral region observed was leaned to the yellow-red spectral region where 
the NaI-D, H$\alpha$, LiI$\: \lambda 6708$ \AA, CaI$\: \lambda 
6718$ \AA, the [OI], [NII] and [SII] nebular lines are, as well as the TiO bands. 
Typically we obtained spectrograms with S/N ratio $\ge 30$ (i.e. $1\sigma \ge 33$ 
m\AA ~precision in the measured line equivalent widths). Stars G and G1 (see Figure~2) 
seemed to be nebulous in the off-set guider monitor with an integration time of 
15 s and were included in the present survey.

A second run was allotted to the program in February 2008 with the SPMO 84 cm telescope 
(f/15). The B\&Ch spectrograph (model 26767) was attached to its Cassegrain focus, 
provided with a Bausch \& Lomb diffraction grid (ruled to 400 lines mm$^{-1}$ and 
blazed at $9^\circ 44'$ to optimize first-order observations). The CCD SITE \#3 and its 
respective housing were used to register the spectrograms. The spectrograph's slit had 
discrete width settings: we used the 80 $\mu$m (1\farcs31 sky-projected) slit width, 
since it met better our requirements. The seeing was measured from the (guided) FWHM 
intensity tracings and was $\lesssim 1$\farcs5, normally 1\farcs0. The grid angle was set 
at $ 7^\circ 45'$ to embrace the spectral range of the 2006 observing run at the 2.1 m 
telescope. The same CuHeNeAr arc lamp was used to calibrate in wavelength the dispersion 
axis. The resulting dispersion and resolution at H$\alpha$ were 3.07 \AA/pixel and 
6.1 \AA, respectively. The same alignment issue of the detector we had in the previous 
run at the 2.1 m telescope persisted, combined with more serious alignment problems of the 
dewar with spectrograph. As a result of this, the spectrogram tracings flared for shorter 
wavelenghts, making them useless for our purposes at wavelenghts  $\lambda \lesssim 5600$\AA . 
The sky-projected pixel size was 0\farcs393 at H$\alpha$. S/N 
ratio of the spectrograms was typically $\gtrsim 15$ (i.e. $1\sigma \gtrsim 75$ m\AA 
~precision in the line equivalent width measurements), about  10  for the weaker 
stars. In the 2008 run the observations were near their limit of detection in a 
reasonable integration time ( $\sim 1800$ sec) for a useful spectrogram.  

\subsection{Data reduction}  
\label{sec: red}
The reduction of the spectrograms was done following standard procedures with
IRAF\footnote{Image Reduction and Analysis Facility by NOAO, operated by AURA, 
Inc. (NSF-USA).} software package (Massey et al. 1992). 
The resulting 
spectrograms were normalized for spectral classification purposes. Once a broad 
classification was estimated, to obtain 
the finer spectral types of the program stars, a cross-comparison of the problem 
spectrograms with those of Jacob et al. (1984) 
and the SPMO libraries of 
spectrograms of stars with known MK spectral types was carried out, and when 
necessary, assisted by the Moore (1945) 
multiplet table for line identification purposes (see also Chavarr\'{\i}a-K et al. 1979). 
The TiO band heads were also considered, particularly for the noisier spectrograms 
of the cooler stars. Except for the Balmer lines and the NaI-D and CaI resonance lines, 
the ratio of different lines for an even finer effective stellar 
temperature or luminosity determination were senseless because of the noise of the 
spectrograms (see Figures 3 to 5). The resulting spectral types of program stars are 
given in Table~1, column 4. The brighter and weaker stars had 
uncertainties in their classification of about 1 and 2 -- 2.5 subclasses, respectively, and 
their assumed luminosity classes are given in parenthesis (see also $\S6.1$). 

%
\section{Spectroscopic results}  
\label{sec:spec.res}

Figures 3, 4 \& 5 display normalized intensity spectrograms of the program 
stars, shifted vertically by arbitrary amounts and where strong 
H$_\alpha$ lines were truncated to enhance weaker photospheric 
lines and for clarity purposes. Table~1 summarizes the resulting MK spectral 
classification and the nature of the program stars. The nature of a 
given program star was ascribed taking into account its morphology, 
its MK spectral class, line emission spectra and the intensity of the 
LiI$ \lambda\: 6707$ resonance absorption compared with the CaI$\lambda 
\: 6717$ absorption line (i.e. the W(LiI)/W(CaI) line ratio). 
From medium and high resolution spectra of a representative sample of T Tauri 
stars we call the attention to the reader that, Wichmann et al. (1999) conclude that 
medium-resolution spectroscopy ($\approx 4$ \AA) is a robust tool to single out stars 
with high lithium content.None of their Li-rich stars observed with medium-resolution 
proved wrongly classified with the follow up high dispersion data.

The spectral survey of the IRAS06548-0815 region revealed previously 
unknown pre-main sequence stars: we found two Herbig-emission stars, 
two classic T Tauri and four weak-line T Tauri stars,  and two probable 
lithium-rich stars. Additionally we found that IRAS06547-0810 is excited 
by a single B1~V star associated with reflection nebulosity, and that 
NJ065703.0-081421 and NJ065714.1-081016 are an A6~Vp star and a B3/B4 V 
star, respectively. For other relevant data of the program stars, consult 
Table~1.   

\begin{deluxetable}{lcccrccccl} 
\tabletypesize{\scriptsize}
\rotate
\tablecaption{Spectroscopic results\label{tbl-1}}
\tablewidth{0pt}
\tablehead{
\colhead{star} & \colhead{$\alpha_{2000}$} & \colhead{$\delta_{2000}$} & \colhead{Sp.T.} & \colhead{W(H$\alpha$)~\AA} &
\colhead{W(LiI)~\AA} & \colhead{W(CaI) ~\AA} & \colhead{run} &
\colhead{membership\tablenotemark{a}} & \colhead{remarks} 
          }
\startdata
  A & \hspace*{-4em}06$^{h}$ 57$^m $ 09$\fs$20 & -08$\degr$ 20$\arcmin$ 42$\farcs$2 & F5 V    &~~~8.9 & $<0.05$ & 0.05\tablenotemark{b} & 2.1 m & nm & frgrd. \\ 
  B & 14$\fs$35 &~~~ ~~ 19$\arcmin$ 56$\farcs3$ & B2 (V)p   &-~79.8 & ~--~    & ~--~  & 2.1 m & m  & Herbig Ae/Be    \\ 
  C & 14$\fs58$ &~~~ ~~ 19$\arcmin$ 51$\farcs2$ & K5/K6 (V) &-106.7 & ~1.5    & ~1.4  & 84 cm & m  & CTTS            \\ 
  D & 14$\fs58$ &~~~ ~~ 20$\arcmin$ 04$\farcs8$ & B2/B3 Vp&~~~7.2 & ~--~    & ~--~  & 2.1 m & m  & nebulous,       \\ 
  E & 15$\fs16$ &~~~ ~~ 19$\arcmin$ 56$\farcs3$ & F9 V    &~~38.7 & ~3.1    & 0.05  & 84 cm & m? & WTTS            \\ 
  F & 15$\fs$39 &~~~ ~~ 20$\arcmin$ 04$\farcs8$ & G0\tablenotemark{d} V    &-~14.9 & ~0.4    & ~0.9  & 2.1 m & m? & Li-rich:\tablenotemark{b}        \\ 
  G & 15$\fs$39 &~~~ ~~ 23$\arcmin$ 59$\farcs4$ & A8/A9 V &-~~8.5 & ~~~  & ~--~  & 2.1 m & dcm & Herbig Ae/Be  \\ 
  H & 19$\fs$39 &~~~ ~~ 20$\arcmin$ 11$\farcs6$ & K1 (V)     &-134.6 & ~0.2    & ~0.2  & 2.1 m & m  & CTTS            \\ 
 H1 & 18$\fs$71 &~~~ ~~ 20$\arcmin$ 13$\farcs3$ & K2/K3 (V) &~~~4.8 & ~0.3    & ~0.2  & 2.1 m & m  & WTTS            \\ 
  i & 11$\fs$26 &~~~ ~~ 19$\arcmin$ 42$\farcs7$ & F7 V    &~~~5.0 & ~--~    & ~--~  & 84 cm & nm & field star(?)      \\ 
  l & 12$\fs$18 &~~~ ~~ 21$\arcmin$ 23$\farcs0$ & K4 (V)    &~~~7.5 & ~0.6    & ~2.2  & 84 cm & m? & Li-rich:\tablenotemark{b}         \\ 
  J & 19$\fs$17 &~~~ ~~ 20$\arcmin$ 28$\farcs6$ & K5 (V)    &~~~2.1 & ~0.3    & ~0.3  & 2.1 m & m  & WTTS            \\ 
  j & 19$\fs$85 &~~~ ~~ 19$\arcmin$ 58$\farcs0$ &  ~--\tablenotemark{c}~   &~--~   & ~--~    & ~--~  & 84 cm & ?  &noisy, NaI D abs present.\\ 
  k & 14$\fs$81 &~~~ ~~ 18$\arcmin$ 56$\farcs7$ & G7/G8 (V) &~~~3.5 & ~1.0    & 0.05\tablenotemark{b} & 84 cm & m  & WTTS            \\ 
    &           &                         &         &       &         &       &       &    &                 \\ \tableline 
    &           &                         &         &       &         &       &       &    &                 \\ 
 G1 & 24$\fs$49 &~~~ ~~ 24$\arcmin$ 07$\farcs9$ & A8 V    &~~10.3 & ~0.1    & ~0.2  & 2.1 m & nm & field star      \\ 
 IRAS06547-0810 
    & 08$\fs$04 &~~~ ~~ 14$\arcmin$ 25$\farcs0$ & B1 V    &~~~3.0 & ~--~    & ~--~  & 2.1 m & cm &                 \\ 
 NJ065703.0-081421
    & 03$\fs$60 &~~~ ~~ 14$\arcmin$ 15$\farcs5$ & A7 Vp   &~~~7.8 & ~--~    & ~--~  & 2.1 m & cm & abs. 5848\AA?  \\
 NJ065714.1-081016 
    & 14$\fs$50 &~~~ ~~ 10$\arcmin$ 15$\farcs5$ & B3 Ve   &~~12.8 &~--~     & ~--~  & 2.1 m & cm? & em. 6375\AA?  \\ 
\enddata 
\tablenotetext{a}{m = member of IRAS06548-0815 cluster; nm = not member of the dark cloud and/or trapezium; dcm = member of the dark cloud} 
\tablenotetext{b}{Uncertain, the line is at about the continuum noise level} 
\tablenotetext{c}{late K  or later type star.}   
\tablenotetext{d}{uncertain, the spectrogram lacks key lines for its spectral type}  
\end{deluxetable}  

\begin{figure}
 \includegraphics[width=1.0\columnwidth]{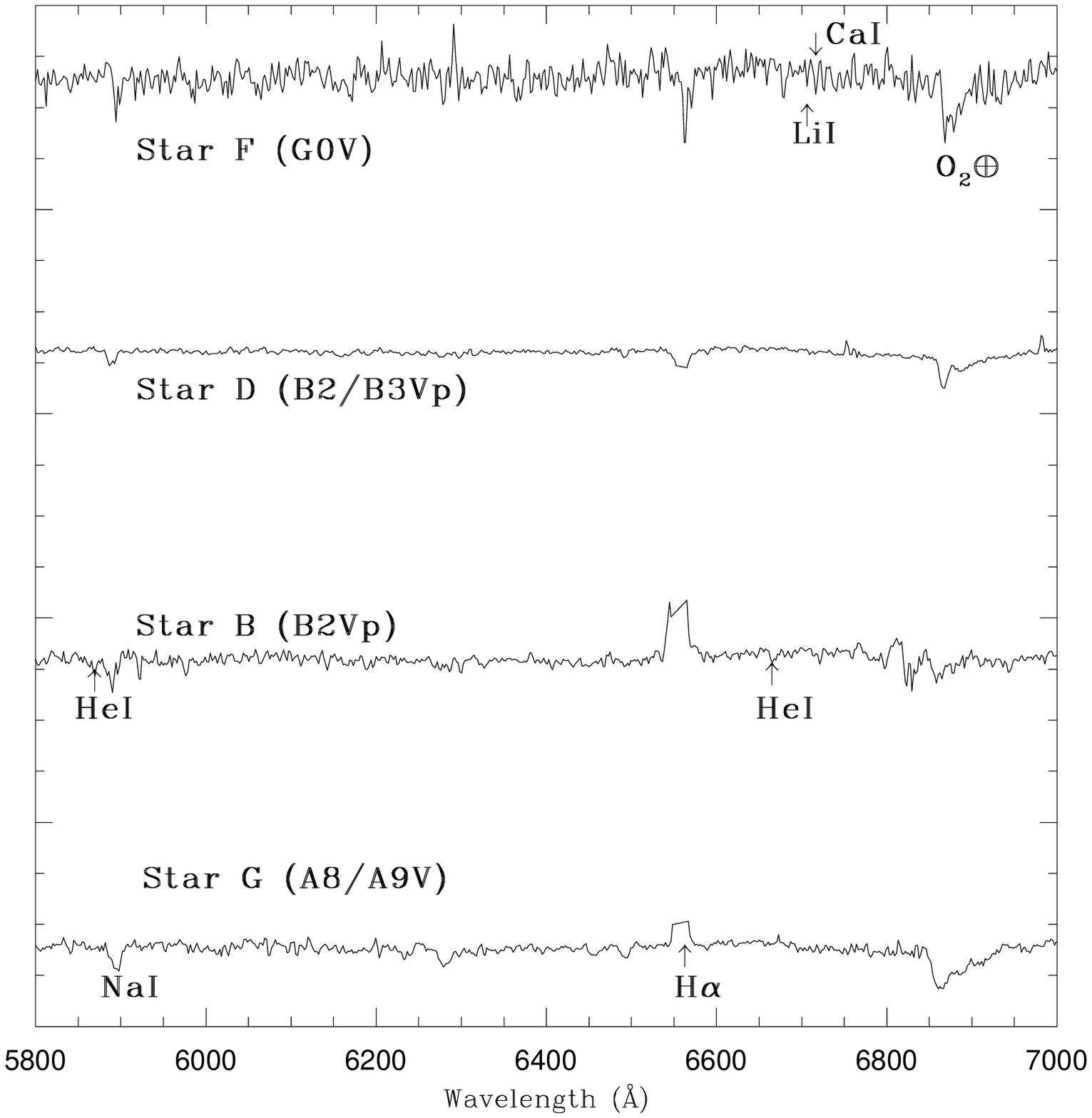}
  \caption{Normalized intensity tracings of the trapezium stars
       B, D, F and the dark cloud member star G, shifted in the intensity 
       axis by an arbitrary amount (at the same scale factor) for display 
       purposes. The strong H$\alpha$ lines of stars B, D and G spectrograms 
       were truncated to enhance the weaker lines (see Table~1 for their 
       equivalent widths). The spectrograms were taken with the 2.1 m 
       telescope. Star designations as in Figures 1 \& 2.
          }
  \label{fig3}
\end{figure} 

%
\begin{figure}
  \includegraphics[width=1.0\columnwidth]{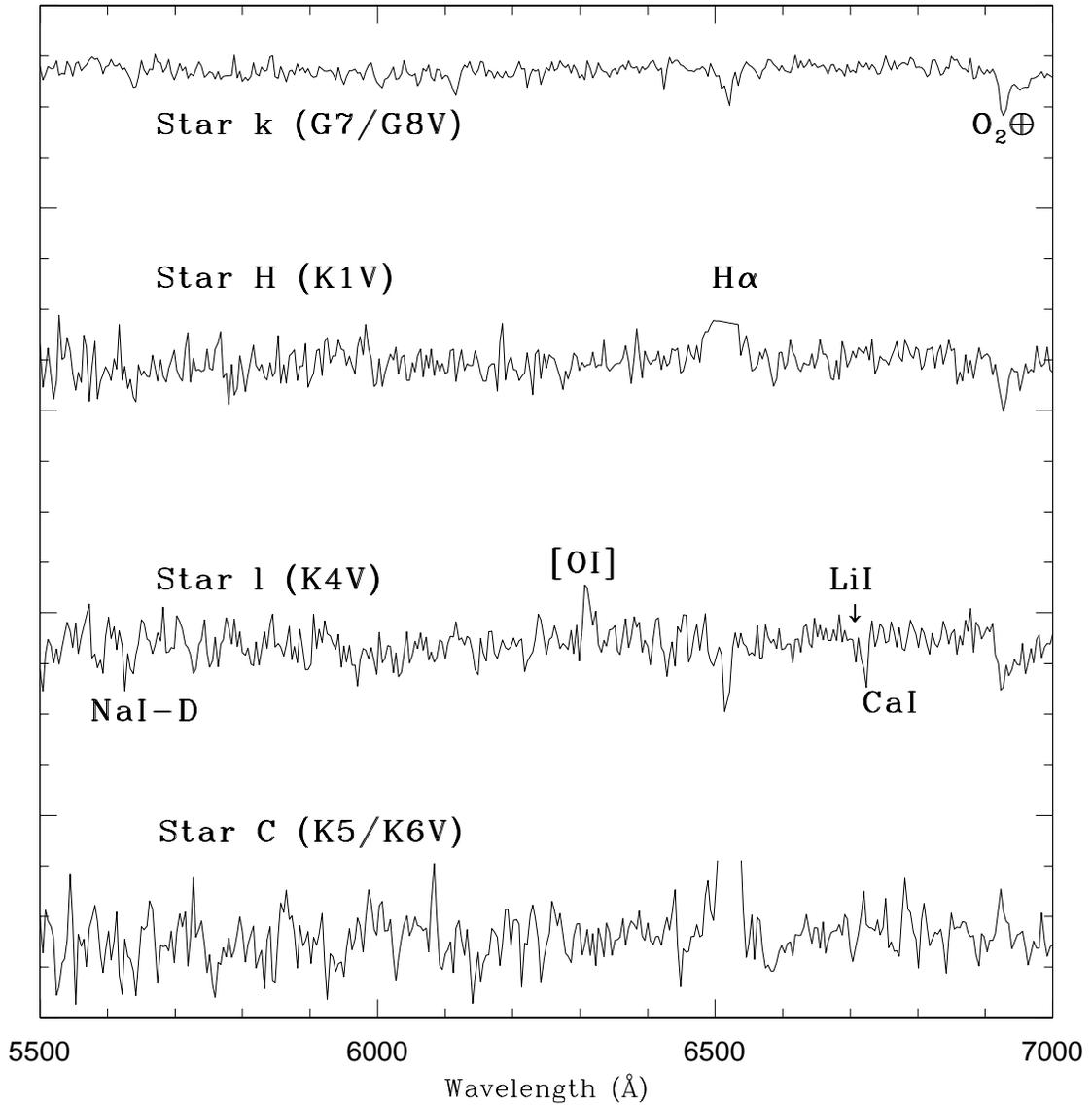}
  \caption{Normalized intensity tracings of the trapezium stars C, H, k and l taken 
       with the 0.84 m telescope, shifted in the intensity axis by an arbitrary 
       amount for display purposes. Stars H, k, l are on the same intensity scale, 
       C is amplified by a factor of two.   
       The strong H$\alpha$ emission lines of stars C and H spectrograms 
       were truncated to enhance the weaker lines.
       Star designations as in Figures 1 \& 2.  }
  \label{fig4}
\end{figure} 

%
\begin{figure}[t]
  \includegraphics[width=1.0\columnwidth]{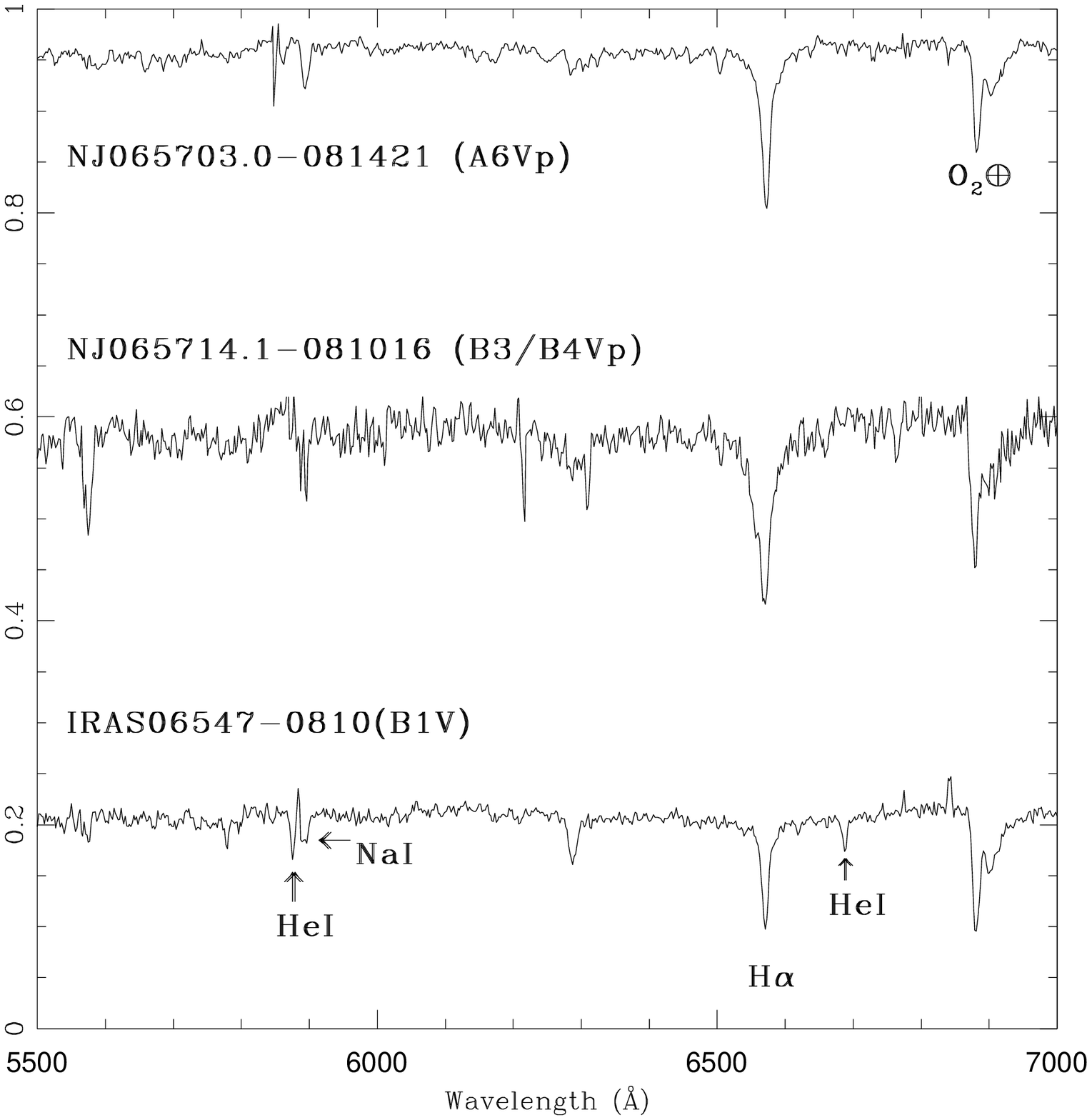}
  \caption{Normalized intensity tracings of stars NJ065703.0-081421, NJ065714.1-081016 
           and IRAS06547-0810 taken with the 2.1 m telescope and displayed at the 
           same intensity scale factor. Star designations as in Figure 1 .}
  \label{fig5}
\end{figure} 

%
\section{Distance to LDN 1655} 
\label{sec:distances} 
\objectname{LDN~1655} is in the proximity of the ``Southern Filament'' CO cloud (Maddalena et al. 1986, 
see also L\'opez et al. 1988), 
located at about $3^\circ$ north of the Southern Filament and $2^\circ$ west of \object{NGC2316}, i.e. 
$\gtrsim 57$ pc and $\gtrsim 38$ pc, respectively, assuming a distance d$_{LDN~1655} = 1.1$ kpc to the 
cloud. 
The Southern Fillament is a continuous bridge of CO emission joining the molecular cloud 
Mon~R2 with the CMa~OB1 association, as the CO velocities varies smoothly between these two 
regions (Maddalena 1986). 
On the one hand, based on the star-count method, these authors 
obtained a ``crude distance estimate'' of $1000 \pm 150 \;$pc to the 
``Cross-Bones'' CO cloud, located in the Southern Filament and L\'opez et al. (1988) 
cite a kinematic distance  of 1.1 kpc to the 
nearby galactic cluster NGC2316 (i.e. LDN~1654). But both methods above can have large inaccurancies. 
On the other hand, from photometric observations of the stars exciting several reflection nebulae 
and hence presumably associated with the \objectname{Mon~R2} cloud, the \objectname{Southern Filament} 
and westward of \objectname{CMa~OB1}, the (more accurate) photometric distances  of 830 pc, 912--1060 
pc and 1200 pc,respectively, were found by different authors (van den Berg 1966; Racine 1968; Herb \& 
Racine 1976; Eggen 1978; Clari\'a 1974 a \&b).  
Since the CO radial velocities vary gently from one end to the other of the above regions, 
interpolating these photometric distances with the corresponding CO velocity of the ``Cross 
Bones´´ CO cloud given by Maddalena et al. (1986),  
we estimate here a distance of $1.1\pm 0.1$ kpc to the Southern Filament/Cross-Bones cloud. 
Moreover,  the Southern Filament extends about 230 pc along the line of sight and about 300 
pc along the plane of the sky (Maddalena et al. 1986). 
However, by inspecting the surface stellar density around the region on the Palomar sky survey prints 
and considering the non-detection of CO between LDN~1655 and the Southern Filament or NGC2316,
LDN~1655 may not belong to the Southern Filament complex at all. 

Considering that the R- or NJ-stars illuminate their immediate surrounding dust thus indicating 
us that they (most probably) are physically associated to the dust cloud LDN1655, one can estimate 
the dark cloud's distance, if the appropiate data of the associated stars is known (i.e. photometry 
and MK spectral types). Here we give the spectral types of five R-stars associated with LDN~1655, 
namely NJ065703.0-081421, NJ065704.1-081016, IRAS06547-0810 and stars D \& E of IRAS06548-0815 
(cf. Figure~2, Table~1), which combined with their {\sl 2MASS} near-infrared photometry allows us 
to obiain the (photometric) distance to the dark cloud, which is more accurate than any kinematic 
estimate. 

According to their spectral types given in Table~1, we proceeded to determine the average total 
interstellar absorption in the K-band $<A_K>$ that results from the color excesses E(J-K), E(J-H) 
and E(H-K) of the individual stars (c.f. Table~3). The color excess in a given color $(\lambda_1 - 
\lambda_2)$ is given by the relation 
\begin{equation} 
E(\lambda_1 -\lambda_2) = (\lambda_1 -\lambda_2) - (\lambda_1 -\lambda_2)_{i}
\end{equation} 
were the subscript $i$ distinguishes the intrinsic from the observed color. The $\lambda$'s stand 
for the magnitudes of the passbands J, H,and K $(\lambda_1 < \lambda_2)$. The  intrinsic colors of 
the stars and the interstellar extinction law we adopted are those given by Koornneff (1983) 
and Mathis (1990), respectively (see also Chavarr\'{\i}a-K et al. 1987). 
They were used to deredden the (observed) K magnitudes of the R-stars correspondingly with the expression 
\begin{equation}
K_{12\circ} = K_s - \frac{A_K}{E(\lambda_1-\lambda_2)} \times E(\lambda_1-\lambda_2)_{obs} 
\end{equation}
where $\frac{A_K}{E(\lambda_1-\lambda_2)}$ is given in Table~2 and $E(\lambda_1-\lambda_2)_{obs}$ 
is the observed color excess of the program star in question.
\begin{table}
\caption{Adopted i.s. extinction law} 
\vspace*{1mm} 

\begin{tabular}{cccc} 
\tableline \tableline  
$\frac{A_K}{E(H-K)}$&$\frac{A_K}{E(J-H)}$&$\frac{(A_K)}{E(J-K)}$ & $\frac{A_K}{A_J}$\\ \tableline
        1.579       &         1.016      &          0.618        &   0.382    \\ 
\tableline 
\end{tabular} 
\end{table}   

With help of the i.s. extinction law of Table~2 and the color excesses of the 
program stars given in Tables~3 \& 4 , the average unreddened $<K_{s\circ}>$ 
magnitudes of the program stars are readily calculated with the expression 
\begin{equation}
 <K_{s\circ}> = \frac{1}{3} \times (K_{JH\circ} + K_{HK\circ} + K_{JK\circ})
\end{equation} 
for all stars of Tables~3 \& 4, except for the classic T Tauri star IRAS06548-0815~C 
because of its large infrared excess in K: its J magnitude was corrected for 
interstellar absorption using Eq.~2 and the (observed) E(J-H) color excess and with 
help of its suitable intrinsic color, we derived the unreddened visual magnitude of 
this star. 

One can derive the distance to a star from its dereddened infrared apparent magnitude  
\mbox{$<K_{so}>$}  and the absolute K magnitude $M_K$ corresponding to its MK spectral 
type directly or calculate its deredden visual magnitude $V_o$  from its $<K_{so}>$ with 
help of the intrinsic colors corresponding to its MK spectral type and, with its absolute 
visual magnitude $M_V$, estimate its distance from the observer. Since the 
former method intrinsically depends from (the calibrations of) the latter, we have chosen 
here this last procedure. The unreddened visual magnitudes $V_{\circ}$ of 
the program stars were obtained from their (average) dereddened $<K_{s\circ}>$ magnitudes 
by adding the intrinsic color $(V-K)_{i}$ 
corresponding to their MK spectral type  (Koornneeff 1983). From these results, together 
with the Schmidt-Kaler (1982, SK82 hereafter) absolute magnitude M$_V$ vs. 
MK spectral type calibration for main sequence stars, we estimated the 
distance moduli of the individual stars. The distance to the dust cloud was 
obtained by averaging the distances of the individual objects given in Table~3. 
This method has the advantage that for a (relatively) small uncertainty in 
the near-infrared color excesses and hence in the dereddened K~magnitude derived 
from Equation (2) remains small in the corresponding $V_{\circ}$, regardless 
of the adopted interstellar extinction law\footnote{for $\lambda \geq 0.9 
\micron$ the i.s.  extinction law is the same for diffuse dust $R_V = A_V/E(B-V)=3.1$ 
and for outer dark cloud environment $R_V\simeq 5$.}(Chavarr\'{\i}a-K et al. 1987, Mathis 1990). 

\begin{deluxetable}{lccccccccc}
\tabletypesize{\scriptsize}
\tablecaption{individual distances of the R-stars\tablenotemark{a}} 
\tablehead{ 
\colhead{star}     &  \colhead{K$_s$}   &   \colhead{E(J-H)}       &  \colhead{E(H-K)}   &  
\colhead{E(J-K)}   &   \colhead{$<A_K>$\tablenotemark{b}} &  \colhead{$<K_{s\circ }>$}  &  
\colhead{$V_{\circ}$} & \colhead{$5\,log\,D-5$} &  \colhead{D(pc)} 
           } 
\startdata 
NJ065703.0-081421& 11.80& 0.05 & 0.05 & 0.08 &    0.00    &   11.82        & 12.27 & 10.07 & 1033 \\  
                 &  ~~02& ~~03 & ~~05 & ~~08 &   ~~~02    &   ~~~03        & ~~~04 & ~~~05 & ~103 \\ 
NJ065714.1-081016& 13.51& 0.46 & 0.13 & 0.59 &    0.35    &   13.16        & 12.59 & 14.19 & 6887\tablenotemark{c} \\ 
                 & ~~~04& ~~06 & ~~07 & ~~03 &   ~~~03    &   ~~~05        & ~~~05 & ~~~06 & ~700 \\  
IRAS06547-0810   & ~9.34& 0.74 & 0.35 & 1.09 &    0.66    &   ~8.68        & ~7.92 & 11.12 & 1675 \\ 
                 & ~~~02& ~~03 & ~~05 & ~~02 &    ~~03    &   ~~~04        & ~~~05 & ~~~05 & ~168 \\ 
IRAS06548-0815B  & 10.96& 0.86 & 0.57 & 1.43 &    0.88    &   10.08        & ~9.41 & 11.81 & 2301 \\ 
                 &~~~02 & ~~03 & ~~05 & ~~02 &    ~~03    &   ~~~04        & ~~~04 & ~~~05 & ~230 \\ 
IRAS06548-0815D  & 11.30& 0.71 & 0.40 & 1.10 &    0.67    &   10.63        & 10.01 & 12.01 & 2523 \\ 
                 & ~~~02& ~~03 & ~~04 & ~~03 & ~~   03    &   ~~~04        & ~~~05 & ~~~06 & ~252 \\  \tableline 
                 &     &       &      &      &    &  &\multicolumn{2}{r}{mean of 4 stars}&  $1883\pm 200$     \\ 
\enddata 
\tablenotetext{a}{Formal errors are given in the second row of each star's entrance}
\tablenotetext{b}{$<A_K> = K_s - <K_{s\circ }> $}
\tablenotetext{c}{Not considered in the distance estimate of LDN1655, see text for details}
\end{deluxetable} 

Note that star NJ065714.1.1-081016 either does not belong to the dark cloud (it deviates 
too much from the mean of the 5 stars) or it was misclassified spectroscopically (too early 
a spectral type for its apparent K magnitude), or it is a Herbig Ae/Be star but then it 
lacks key spectral features of pre-main sequence stars. Discarding this star from the rest, 
we obtain a mean distance of $1.9\pm 0.3$ kpc to the cloud (more details in $\S6.2$), supporting 
the nonmembership of LDN1655 to the Southern Filament complex. 
Also noteworthy is that NJ065714.1.1-081016 is the brightest star among several stars in 
its immediacy (located $\approx 33^{''}$ southwest of  NJ065714.1.1-081016), apparently associated 
with nebulosity that seen in the {\sl 2MASS} K$_s$ image conform an arch of weak reddened stars, an indication of another possible (localized) low mass star forming region associated with LDN1655. 

%
\section{IRAS06548-0815 and LDN~1655 selected stars in the HR-diagram}  

Following the procedure outlined in $\S4$, we corrected the {\sl 2MASS} $K_s$  
magnitudes of the program stars for i.s. extinction. Using Koornneff's 
intrinsic colors we derived from $<K_{s\circ}>$, 
their corresponding dereddened visual magnitude $V_\circ$ and, assuming a 
distance D = 1.9 kpc to LDN~1655, their absolute magnitudes    
$ M_{V}\: (= V_\circ - 5\: log D + 5).$  
Using the calibration of  the bolometric correction BC with 
spectral type of SK82 we obtained the absolute bolometric magnitudes 
$M_{*} = M_V + BC) $
of the program stars. The luminosities were derived with the expression 
\begin{equation}
log\:(L_*/L_\odot) = -0.4 \times (M_* - M_\odot) 
\end{equation}  
where  $L_*$ and $L_\odot$ are the stellar and Sun luminosities, and $M_*$ and 
$M_\odot (= 4.64)$ are the stellar and solar absolute bolometric magnitudes, 
respectively. 

\begin{deluxetable}{cccccccc}
\tabletypesize{\scriptsize}
\tablecaption{Crude luminosities of selected bright stars associated with LDN~1655} 
\tablewidth{0pt} 
\tablehead{ 
\colhead{star}       & \colhead{$<A_K>$\tablenotemark{a}}     & \colhead{$<K_{s\circ}>$}     & 
\colhead{$V_{\circ}$}& \colhead{M$_{*}$}     & \colhead{$log\: L_*/L_\odot$}& 
\colhead{ age~[yr] } &\colhead{M$_*/M_\odot$}
} 
\startdata 
 ~A\tablenotemark{b} & ~-0.0f4\tablenotemark{c} &~9.56     &10.57  &-0.89       &  ---          & ---               &  --- \\ 
 B         & ~1.16          &~9.80    &~9.13  &-4.54      & 3.67          & $<5.0\: 10^5$       &  $\simeq$ 8 \\
 C         & ~2.43          &~7.44\tablenotemark{d}&10.44   &~3.63        & 0.41          &$5.0 10^5$            & 1.1  \\
 D         & ~0.95          &10.35    &~9.78  &-3.68       & 3.33         & $<5.0\: 10^5$      &  $\approx$ 7  \\
 ~E\tablenotemark{d}          & ~0.82          &10.26    &11.43  &~0.06       & 1.88          & $5.0\: 10^5$      & 4.2  \\ 
 F         & ~0.06          &12.54    &13.76  &~2.26       & 0.95          & $ 4 \: 10^6$      & 2.1  \\ 
 G         & ~0.80          &~9.51    &10.13  &-1.29       & 2.37          & $4\: 10^5$      & 6  \\ 
 G1        & ~0.53          &13.76    &14.32  &~2.90       & 0.70          & $\approx 3.0\: 10^7$ & 1.7  \\ 
 H         & ~-0.28\tablenotemark{c}&12.62    &14.62  &~2.93       & 0.69          & $2\: 10^6$      & 2.5  \\
 H1        & ~1.00          &11.40    &13.77  &~1.99       & 1.06          & $5.0\: 10^5$      & 3.0  \\ 
 i         & ~0.33          &12.23    &13.62  &~2.15       & 0.85          &  -- & -- \\ 
 l         & ~-0.18\tablenotemark{c}&12.79    &15.54  &~3.67       & 0.39          & $6\: 10^6$      & 2.0 \\
 J         & ~0.18          &12.95    &15.95  &~3.91       & 0.29          & $2.0\: 10^6$      & 1.3  \\ 
 ~j        &    ~           &    ~    &   ~   &      ~     &      ~        &        ~          &   ~   \\ 
 k         & ~-0.02\tablenotemark{c}& 12.28   &13.88  &~2.21       & 0.93          & $2.0\: 10^6$      & 1.4  \\ 
\enddata 
\tablenotetext{a}{Is the arithmetic average of the A$_K$'s that result from the individual stellar color excesses E(J-H), E(H-K) and E(J-K) and 
     the relation summarized in Table~2.} 
\tablenotetext{b}{Foreground star} 
\tablenotetext{c}{$<A_K>$ is assumed zero.} 
\tablenotetext{d}{IR excess in K;  stellar luminosity, age and mass derived fitting the SED to the J magnitude, where 
      the infrared excess is the smallest.} 
\tablenotetext{e}{See Section $\S5$ for details regarding this star} 
\end{deluxetable}

The resulting luminosities are given in Table~4 and the effective temperatures used 
to construct Figure~6 were taken from the $(SpT,log\: T_{eff})$ calibration by SK82 
for main sequence stars. The locations of the program stars in the HR diagram are 
shown in Figure~6. The following results for the program stars were inferred from 
their location in the HR-diagram in combination with the PMS evolutionary tracks of 
Siess et al. (2000) computed on line: 

\begin{enumerate}
\item[~~~i)] Stars IRAS06548-0815 A \& G1 (see Figure~2) are, most likely, foreground  
              objects. 

\item[~~ii)] The CTTS nature of star C derived spectroscopically is congruent with  
             its position in the HR-diagram. Star H occupies the lower bend or  
             transition from convective to a radiative evolutionary path (see Fig~5) 
             and could be a T Cha-like star that mimicks CTTS as well as WTTS 
             alike.

\item[~iii)] Stars H1 and k are spectroscopically WTTS, but star H1 could be a T CHa-like 
             star. The WTT star J occupies the lower bend or transition from 
             a convective to a radiative evolutionary path too. 

\item[~~iv)] The lithium-rich stars F and l evolve radiatively towards the main 
             sequence. 

\item[~~~v)] Considering the uncertainties involved, star i could be a member 
             of the \mbox{IRAS06548-0815} dark cloud if its luminosity class 
             III we suggest here is correct. 

\item[~~vi)] Stars B (mass $\ge 7M_\odot$)  and G (stellar mass $\lesssim 4M_\odot$) are   
             Herbig Ae/Be stars. 

\item[~vii)] IRS1 is most probably an O8 ($ \pm 1$ spectral subclass) star, assuming that it 
             is a main sequence star 
             and hence is the principal exciting star of the IRAS source and of 
             the emission nebulosity detected in the north-northwest side 
             of the visual cluster.  

\item[viii)] Star E occupies the locus of Herbig emission stars in the HR-diagram
             but does not show  spectral features of the class. It probably 
             is a foreground star.

\item[~ix)] The spectrogram of the late type star ~j~ is too noisy to draw any conclusion about its 
            nature. 

\end{enumerate} 

%
\section{Error estimates} 

\subsection{Errors in the MK spectral type classification} 
The main error in the determination of the MK spectral type of a given star lies in the 
signal-to-noise (S/N) ratio of the spectra of both the program stars and comparison 
stars, and in the assigned spectral classification of the program and comparison 
stars. Best achievements result when the two sets of spectra have reasonable 
signal to noise ratios (S/N$ \gtrsim 50$), and about the same spectral dispersion and 
resolution, which is usually not the case. 

 After a first coarse estimate of spectral type of the program star in question that  
depends on the more conspicous lines/features present (or absent) in its spectrum, the 
normilized spectrogram of the program stars were interpolated between two or more 
(normalized) \& similar spectrograms from a library 
of spectrograms of stars with known MK spectral types taken with a similar 
resolution, until a reasonable overall match between the program and comparison stars is 
achieved in temperature  and luminosity classes. We mainly used the Jacoby et al. (1984) 
and the SPMO spectrogram libraries of stars with known MK spectral types to cross correlate 
them with the program stars until an (eyeball) overall best fit of the two was reached. It is 
our experience that following this procedure, 
an uncertainty of one spectral subclass or even less is expected for the brighter stars and about 
two subclasses for the fainter stars  (e.g. Alcal\'a et al. 1996, Chavarr\'{\i}a-K et al. 1979, 
2005, Moreno-Corral et al. 2002, 2006). Because the later spectral type program stars were almost 
at the limit of the 84 cm telescope, and the signal to noise ratio of the resulting spectrograms 
degraded accordingly, an accuracy of 2.5 subclasses in the determination of their spectral class is more realistic.
Spectral types of young stars derived this way by us and in common with other authors 
(e.g. Cohen \& Kuhi 1979; Herbig 1977; Herbig \& Bell 1988) 
compare well within the above given uncertainties. 
On the other hand, regarding the determination of luminosity class, the overall S/N ratio of 
the spectrograms of the program and comparison stars made them unsuitable for a reliable luminosity 
classification in terms of line ratios or the presence of luminosity sensitive photospheric 
lines. An exception are the (profiles of the) Balmer and strong H-like resonance CaI and 
NaI-D lines for early or intermediate and later type stars, respectively. In general, we assumed 
luminosity class V for the program stars (given between parenthesis in Table~1). We justify this 
assumption in terms of the youth of the star members of LDN~1655 and on the results obtained here. 

\subsection{Errors in the estimated distance to the dust cloud LDN~1655}  
The main sources of error in the distance estimate for the dark cloud LDN~1655 are, 
in order of importance,  
\begin{enumerate}
 \item[~~i)] the uncertainties in the assigned spectral types of the associated nebulous NJ- or R-stars. 
\item[~ ii)] the uncertainties in the absolute magnitude  vs. MK spectral type calibration.  
 \item[iii)] the adopted intrinsic stellar colors, 
 \item[~iv)] the asumed reddening law and 
 \item[~~v)] the errors in the {\sl 2MASS} photometry.  
\end{enumerate}

Typical errors in the {\sl 2MASS} infrared magnitudes of the brighter stars are 
$\le 0\fm 03$ (about twice as large for the fainter stars) and we consider them 
negligible when compared to other uncertainty sources involved. Fortunately, the 
interstellar extinction law for wavelengths $ \lambda \geq 0.9\: \mu$m is practically 
the same for any sky region involved (e.g. Chavarr\'{\i}a-K et al. 1987, Mathis 1990, and 
references therein), 
so  we can consider it {\sl universal}, and its contribution to the distance 
uncertainty is small. We expect a final visual magnitude error of the R-stars $\delta_V 
\lesssim 0\fm07$ for stars earlier than F0V and $\lesssim 0\fm25$ for stars later than 
K1V when correcting their $<K_{s\circ}>$ magnitude for reddening and then fitting them 
with the appropriate spectral energy distribution to obtain their average 
unreddened visual magnitude $V_{\circ}$. On the other hand, an uncertainty of (at the most) 
1.5 subclasses, for an average of the R-stars studied here ($\approx$ B2 V), reflects an 
uncertainty of $\delta_{M_V} \simeq 0\fm4$, and the error of the bolometric corrections with 
the stellar temperature  is $\delta_{BC}\sim 0\fm3 $. The above uncertainties add up to 
an expected error $\approx 0\fm4$ in the distance modulus of a single star; however, 
four stars were involved in our distance estimate for the dark cloud, so the final 
uncertainty would be $\simeq 0\fm 3$ in its distance modulus, or an uncertainty of $\pm 
0.26$ kpc for the photometric distance estimate for LDN~1655 of 1.88 kpc derived here, 
in contrast with the earlier estimates of 1.1 kpc with a formal error bar of $\pm 0.1$ kpc.  
           
\section{Discussion and conclusions} 
\label{sec:discuss} 
Assuming a distance of 1.1 kpc, except for stars G, IRAS06548-0815 E and H1, all other 
stars studied here (see Table~1) would be below the zams by 0.28 dex ($\Delta $V$ \simeq 
0_\cdot^m 7$) or more. With $D_{LDN1655} = 1.9 \pm 0.3$ kpc, they fall on 
or above the zams, giving support to the distance derived here. The accuracy of the 
photometric distance estimate given here can be 
significantly improved if the spectral type of NJ065714.1-081016 is redetermined, and more 
R-stars associated with LDN~1655 are observed spectroscopically (i.e. \mbox{IRAS~06550-0805} 
and the two nebulous stars located at $(\alpha,\delta)_{2000} = (06^h56^m47^s_\cdot 6,-08^\circ 14'51'')$ 
and \mbox{$(06^h57^m27^s_\cdot 8,-08^\circ 09'26'')$}, respectively. The stars associated 
with \mbox{IRAS06548-0815} follow the loci of young intermediate and low-mass stars in the 
HR-diagram, confirming their PMS nature. The infrared source 
NIRS~1 most probably excites \mbox{IRAS06548-0815} and its reflection \& emission nebulosities 
(we observed a nebular emission at the NW tip of the cluster). IRAS06548-0815~C is the 
second brightest object in the K band associated with IRAS06548-0815, and is apparently  
a barely resolved binary at this wavelength. Spectroscopy to confirm the MK spectral type 
of \mbox{NJ065704-081402} is recommended. The stellar masses and ages of the program pre-main 
sequence stars given in Table~4 were obtained by eye-fitting their location in the HR-diagram 
using the grid of evolutionary traces computed by Siess et al. (2000) 
and are approximate. Spectroscopy with better S/N ratio and resolution to confirm the MK spectral 
types is recommendend. Spectroscopy of the region about $33^{``}$ south-west of NJ065714.1-081016 
may be rewarding. 

\begin{figure}[t!] 
  \includegraphics[width=1.0 \columnwidth]{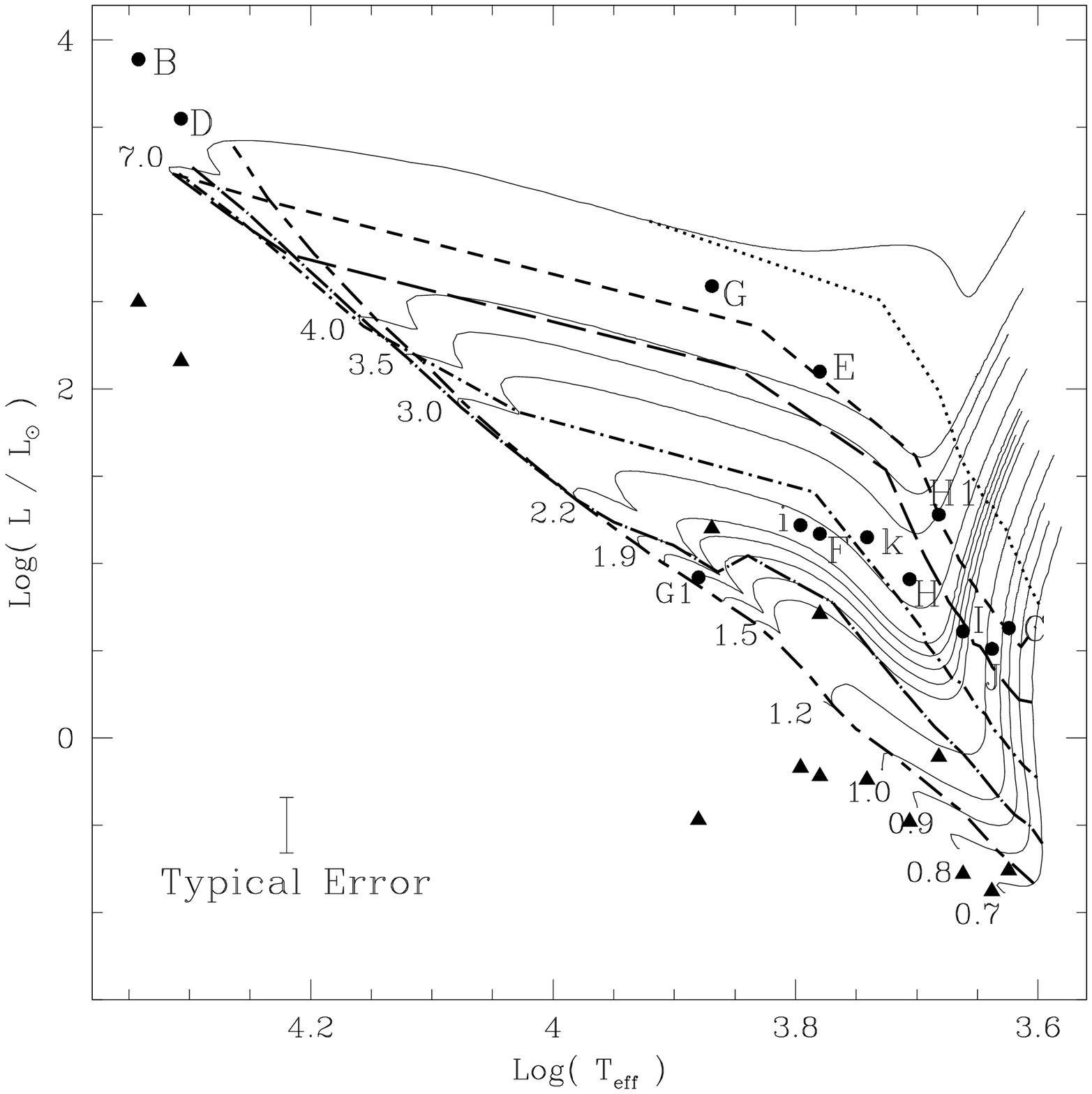}
  \caption{HR-diagram of young stars associated with IRAS06548-0815. The filled circles 
and triangles depict their location in the diagram at $D = 1.9$ and $1.1$ kpc, respectively. 
The grids of evolutionary paths and isochrones for a metallicity Z = 0.02 were taken from 
Siess et al. (2000) 
and were computed on-line in their given WEB site. The isochrone 
ages depicted are as follows: $0.2\times 10^6$ yr = dots; $0.5\times 10^6$ yr = short dashs, 
$1\times 10^6$ yr = long dashs, $3\times 10^6$ yr = dots and short dashs, $10\times 10^6$ yr 
= dots and long dashs, and $30\times 10^6$ yr = short dashs - long dashs. 
          }
  \label{fig6}
\end{figure} 

Considering typical values of $M_\star,\; L_\star$ and $ \;R_\star$ for a given 
contraction phase towards the main sequence of the young stars classified here, 
as derived from the observational data (Cohen \& Kuhi 1979), 
Herbig Ae/Be stars contract with a Helmholtz-Kelvin time scale\footnote{$\tau_{HK}
 =  3.15\; 10^7\: (M_\star/M_\odot)^2 (R_\odot/R_\star) (L_\odot / L_\star)$ yr.} 
$\tau_{HK} < 1 \times 10^6$ yr, CTTS are expected to cover their full convective 
or Hayashi evolutionary phase towards the main-sequence  in a time scale $\tau_{HK}
\approx 1 \times 10^6$ yr, whereas the less active weak-line T Tauri stars (W(H$\alpha) 
< 10\:$\AA) cover their (radiative or Henyey) evolutionary phase towards the main 
sequence with a time-scale $\tau_{HK} \gtrsim 1 \times 10^7$ yr. Low-mass lithium-rich 
stars are young main sequence stars that still have their primal lithium content ($\sim 
3\: 10^7$ yr$ < \tau_{HK} < 2 \times 10^8$ yr). 
Contrary to the times given above, Herbig Ae/Be, CTT, WTT and LiI-rich stars associated 
with IRAS~06548-0815 are dynamially coeval within the contraction/dispersion time of the 
complex (see Section~1 ), a significantly smaller than the shortest Helmholz-Kelvin time 
scale $\tau_{HK}$ of its conforming stars. Although the age estimates of a significant 
portion of the project stars are of about the dynamical time scale of the IRAS06548-0815 knot, 
the evolutionary models predict larger ages fot the rest of them.

Nevertheless, given the time-scales in each evolutionary phase for the pre-main 
sequence and lithium-rich stars, a such admixture of objects constrained in such a small space 
($\phi \approx 0.3\; pc$), IRAS06548-0815 is a remarkable region to study meticulously 
with better instrumentation. 
%
\acknowledgments
We appreciate the support offered by the SPMO mountain staff during the observing runs, in 
particular the night assistants G. Garc\'{\i}a\footnote{Passed away doing his duty in late 
autumn 2010.}, S. Monroy and F. Montalvo. The project 
was partially supported by grant \mbox{400354-5-277757-E} of Mexican Council for Science and 
Technology (CONACYT), and makes use of data products of the Two-Micron All Sky Survey\footnote{
supported by the University of Massachusetts and the Infrared and Processing Analysis 
Center/California Institute of Technology, funded by NASA and NSF (USA).} and of the Centre de 
Donn\'ee Astronomiques de Strasbourg, (CDS, France).  
Our thanks to D. Clark and C. Harris for proofreading an earlier version of the manuscript. 
We are indebt to two anonymous referees for their remarks that significantly helped to improve 
the manuscript.

%


%
\end{document}